\documentclass[12pt]{article}

\def\ppp{\hat{\overrightarrow{\tilde{\pi}}}}

\def\vvv{\hat{\overrightarrow{\tilde{\varphi}}}}

\def\bN{\bf{N}}

\def\ve{\vskip.5em}
\def\k{\kappa}
\def\hq{\hat{\varphi}}

\def\vp{\4varphi}

\def\v{\vskip1em}

\def\Ra{\Rightarrow}

\def\half{\textstyle{\frac{1}{2}}}

\def\H{{\cal H}}

\def\vp{\varphi}

\def\H{{\cal H}}

\def\l{\lambda}

\def\ra{\rightarrow}
\def\tint{{\textstyle\int}}

\def\hp{{\hat\pi}}

\def\d{\partial}
\def\dag{\dagger}

\def\b{\begin{eqnarray*}}  
\def\e{\end{eqnarray*}}    
\def\bn{\begin{eqnarray}}  
\def\en{\end{eqnarray}}   

\def\<{\langle}
\def\>{\rangle}

\def\bk{\mathbf k}

\def\no{\nonumber}

\def\ds{d^s\!x}
\def\k{\kappa}

\def\hk{\hat{\kappa}}
\def\{{\lbrace}
\def\hv{\hat{\varphi}}
\def\d3{d^3\!x}

\def\b{\beta}

\def\tp{\tilde{\pi}}
\def\tk{\tilde{\kappa}}
\def\tv{\tilde{\varphi}}
\def\bN{\bf{N}}
\def\htp{\hat{\tilde{\pi}}}
\def\htk{\hat{\tilde{\kappa}}}
\def\htv{\hat{\tilde{\varphi}}}

\def\htp{\hat{\tilde{\pi}}}
\def\htk{\hat{\tilde{\kappa}}}
\def\htv{\hat{\tilde{\varphi}}}

\def\}{\rbrace}

\begin{document}

\title{ A Valid Quantization of 
The Particle in a Box Field Theory, 
and Well Beyond}
\author{John R. Klauder\footnote{klauder@ufl.edu} 
\\Department of Physics and Department of Mathematics  \\ 
University of Florida,   
Gainesville, FL 32611-8440}
\date{}
\let\frak\cal

\maketitle 
\begin{abstract} 
The usual particle in a box is turned into a field theory, and its behavior is examined using canonical and affine quantizations. The result leads to a valid affine quantization of the  particle in a box field theory, which points toward further valid quantizations of more realistic field theory models. 
\end{abstract}
\section{Introduction}
A recent article by the author \cite{k1} considered the half-harmonic oscillator field theory, using both canonical and affine  quantization procedures.
The results of that study led to a valid quantization using affine quantization. In this paper, we follow a similar path to turn the particle  in a box model \cite{www} into a field theory, and in this paper, we exploit a novel version of a `quantum field theory'.  Hereafter, we often will use CQ as an abbreviation for canonical quantization along with AQ as an abbreviation for affine quantization.

Such a simple model can also shed information on the family of traditional covariant   quantum field theories. Indeed, we will illustrate a novel approach to a set of common covariant quantum field theory known as  $\vp^p_n$.

\section{Establishing the Classical Story}
The classical Hamiltonian is chosen as $H=p^2$, and the range of coordinates is $-b<q<b$, where $0<b<\infty$. Evidently, the particle must bounce off the two `walls', located at $q=\pm b$, as it travels back and forth.

\subsection{Multiple identical particles, finite  and, formally, infinite}

The next step is to collect many such terms which are completely independent of one another. Now the classical Hamiltonian is $H=\Sigma_{k=1}^N \;p^2_k$, while the coordinate space becomes $-b<q_k<b$ for every $k\in (1, 2, 3,..., N)$. 

Although nothing is infinitely  substantive, even the number of atoms, we prepare our Hamiltonian as done traditionally, to become a spatial field. Since we do not accept this route later, we temporarily  introduce that of a single coordinate $y$ (using $a$ to become $0$ and $N\ra \infty$), and demand that $Na=100$. Now, we pass to that limit by
 \bn  H_N= \;\Sigma_{k=1}^N\;\;p^2_k\;a\Ra Na=100
 \;,\:\;   H_{N\;\lim_{a\ra0}} \Ra H_c=\tint_0^{100} \pi(y)^2\;dy\;,\en
 while we now have $-b<\vp(y)<b$ for all $0<y<100$.
 The equation of motion for this system involves $\pi(y,t)=\dot{\vp}(y,t)=\pm A(y)$, hence $\vp(y,t)=\pm A(y)\,t + B(y)$, restricted so that $|\vp(y,t)|<b$. The field value, $\vp(y, t)$, oscillates back   and  forth, bouncing off each of the two `walls' that are located so that  $|\vp(y)|= b$.\footnote{Later, in the grand summation, we lead to several dimensions simply by changing $ k\Ra\bk=(k_1, k_2,..., k_s) $. }
 
 This can be good math, but it is not a very good match to physics. This statement becomes more evident in the following section.
 
 \subsection{A path integration  issue}
 An important  part of canonical quantization (CQ) involves path integrations of the classical Hamiltonian in which terms like $K$ (named for the kinetic term)  \bn 
 K= \tint_0^T \{\tint_0^{100} \pi(y,t)^2 \;dy \}
  \;dt \;<\infty \;, \en 
  in which there are two time limits, i.e., $ \pi(y,0)$ and $\pi(y,T)$, both of which are fixed. It follows that the range of paths can lead to `integrable infinities', such as $\pi(y)^2= R(y)/|y|^{2/3}$, for a harmless $R(y)>0$, or for  $\pi(y, t)^2= W(y, t)/|T/2 -t|^{2/3}$, also for a harmless factor $W(y, t)>0$. Such issues complicate conventual quantum field theories using CQ, and now, they  can also complicate this `toy model'. 
  
  We continue dealing with integrations and potentially infinite densities,  aiming  toward alternative formulations. 
  
  \subsection{A procedure to eliminate `integrable infinities'}
  We start by repeating that $\pi(y)^2$ can reach infinity in value. To eliminate this possibility  we introduce the dilation field, $\k(y)\equiv \pi(y)\,\vp(y)$ along with $\vp(y)\neq0$. We deliberately remove $\vp(y)=0$ so that
  $\k(y)$ should not be forced by $\vp(y)=0$ to be zero because that value should be reserved for $\pi(y)=0$, which, after all, is really $\dot{\vp}(x,t)$. 
  Likewise, we require that $|\pi(y)|+|\vp(y)|<\infty$  --- because, any one of them being infinite forces another factor to be infinite as well, and then the third one would be helpless  ---  which insures that  $|\k(y)|<\infty$ as well. Since  $\pi(y,t)=\dot{\vp}(y,t)$, it is acceptable to let $0\leq [\;|\pi(y,t)|+|\k(y,t)|\;]<\infty$, along with    $0<|\vp(y,t)|<\infty$. The fact that $|\dot{\vp}(y,t)|<\infty$ ensures that $\vp(x,t)$ is continuous in time.
  
  Summarizing, the properties outlined above, point to   $\pi(y)^2=\k(y)^2/\vp(y)^2$, which  requires that since $|\k(y)|<\infty$ and $0<|\vp(y)|<\infty$, and now, it follows that $|\pi(y)|<\infty$ ensuring that there are no `integrable infinities'. 
  
  This feature can also apply to discrete summation expressions, prior to passing to their infinite expression for a genuine integration, which then 
  forbids  any potential `integrable infinites' from even being created. 
  
  Still dealing with integrations, we can  choose our classical Hamiltonian field,  using affine variables $\k(x)$ and $\vp(x)\neq 0$, as $H'= \tint_0^{100}\;k(y)^2/\vp(y)^2\;dy$, as well as  $K'=\tint_0^T\{ \tint_0^{100}[\k(y,t)^2/\vp(y,t)^2]\;dy\}\;dt$. This formulation avoids any `integrable infinities' 
  thanks to the rules that,    $0\leq |\k(x,t)|<\infty $ and $0<|\vp,x,t)|<\infty$, which then   ensures  that $0\leq|\pi(x,t)|<\infty$, and all three functions are  related correctly.
  
  \section{Preparations for  The Particle in a Box \\`Quantum Field Theory'} 
  \subsection{A  conventional CQ approach  to quantization}
  The usual field operators are $\hp(x)$ and $\hv(x)$ where $x=(x_1, x_2,...,x_s)$. The standard   commutation term is $[\hv(x), \hp(y)]= i\hbar \,\delta(x-y)1\!\!1$, which essentially  leads to 
  $ \delta \;\vp(y)/\delta \vp(x)= \delta(x-y)$. The later term represents Dirac's $\delta(x)$, which is zero if $x\neq0$ and infinity  if $x=0$, such that $\tint_{x<0}^{x>0} \delta(x)\;dx =1$. This is   formally accepted, but it does not fit good math or not seem to fit good physics either. This is because $ \lim_{(y\ra x
  )} \delta\;\vp(y)/\delta \vp(x)= 0$ while $\delta \;\vp(x)/\delta\vp(x)= \infty$. Not only is the second  limit 
  not zero, it is infinity. Nature, as represented by  physics, does not need to accept discontinuous limits, and should be far more careful with infinities. 
  
  The foregoing story was  included to stress that nature is composed of atoms not of mathematical points of zero size. After all, we build walls with bricks not points. A deck of cards, which is roughly 1cm tall, and consists of roughly 50 cards, is not composed of infinitely many sheets of zero thickness, etc.
  
  We  continue with an effort to make field quantization more physical and far less infinite.
  
  \subsection{A procedure to simulate quantum physics better}
  To begin with, every integration technically starts with the limit of a grand summation, such as
  $\Sigma_{n=1}^N f_n\;a\;\Ra_{a\ra0} \,\Ra\;\tint_A^B f(x) \;dx$ provided that  $Na =B-A$. For the vast amount of useful integrations, stopping at $a=10^{-25}$ and $N=(B-A)\,10^{25}$ would be quite sufficient. 
  Let's introduce a similar game with quantum field operators. 
  
  We begin by introducing the symbol $\tau=a^s>0$ as a substitute for $\ds$, and $\sum_{\bk}^{\bN}$, where $\bk=(...,\bk_1, \bk_2, \bk_3,...)^s$ and ${\bN} =N^s$.
 We approximate $a^s$ to be an $s$-dimensional, 
  `atom sized block', with a `volume', say $a^s =10^{-s\times 10}$ meters$^s$, with `atom blocks' that  stack well together. 
  
  Since conventual integration relies on an infinite  limit of a vast number of tiny  terms,
   there is no reason not to adapt that procedure for quantum theory,  especially for models that are  composed of completely independent terms, like our present model, i.e., 
    ``The particle in a box, field theory''. In that case, we promote each term in the classical Hamiltonian. Using CQ, we are led 
    to the quantum Hamiltonian, given by $\H =\sum_{\bk}^{\bN}\;\htp(\bk)^2\; \tau\;$, where we have renamed our `discrete  atoms', as    
  $P_\bk=\tp(\bk)$ and $Q_\bk=\tv(\bk)$. 
    The labels of 
  the quantum variables are now used to become labels of  a `simulated  field' (hence the tilde), not by infinitely many  `points', but by a uniform stack of  `atom blocks'.\footnote{There are useful procedures that admit coordinate math-like points of quantum fields in which rescaling of the quantum operators is accepted. An example of just that procedure can be seen in \cite{ k1}. However, in the present paper, we choose a different approach altogether.}
    
    \section{An AQ Formulation of \\The Particle in a Box \\`Quantum Field Theory'}
  The classical Hamiltonian of multiple particles is still 
  $ H=\sum_{\bk}^{\bN}  \tp(\bk)^2\;\tau$.
  However, now using AQ, for which $\tk(\bk)=\tp(\bk)\,(b^2-\tv(\bk)^2)$, it follows that  \bn \htk(\bk)=[\htp(\bk)^\dag\, (b^2-\htv(\bk)^2) + (b^2-\htv(\bk)^2) \,\htp(\bk)]/2\;, \en so that 
   \bn &&\hskip-1.4em \H' =\Sigma_{\bk}^{\bN}\,\{\;\htk(\bk)\;(b^2-\htv(\bk)^2)^{-2})\;\htk(\bk)\;\}\;\tau \no\\ &&
   =\Sigma_{\bk}^{\bN} \,\{\; \htp(\bk)^2 +
 \hbar^2[2\htv(\bk)^2 + b^2]/[\,b^2-\htv(\bk)^2\,]^2\;\} \:\tau \;.\en 
 This special $\hbar$-term has been taken from the dilation variation study in \cite{killer-1}.
 
 In these expressions we choose $\tau=10^{-s\times 10}$, which pertains to the  size of typical atoms in meters, while $\bN\times 10^{+s\times 10}$ to fulfill an  
 approximate value of the integral, in which we  let the $\bN\times\tau=s\times 100$ while $\tau\ra0$. 
 However, we do {\bf NOT} do that, but instead we leave the  `grand summation' in effect for the simple reason  that it  carries  much more physics than the final  integration does.
 
  Since we had numerous, identical, toy models of  the particle in a box, their eigenfunctions are identical at every `block',  such as at any $\bk$.
  In that case, eigenfunctions of our grand summation have the form
   \bn \Psi_n(f) =\Sigma_{\bk}^{\bN} f(\bk)
   \; \psi_n(\bk)\;, \en
 where we require that $\Sigma_{\bk}^{\bN} 
 \;|f(\bk)|^2\; \tau<\infty$, and where $n=0,1,2,3,...$ denotes the ordering of the individual eigenvalues by their increasing magnitude.   At the present  time, the basic AQ quantization of a particle in a box, i.e., 
 \bn \hbar^2\, [-(d^2/dx^2) + [2x^2+b^2]/[b^2-x^2]^2\;\psi_n(x) = E_n\;\psi_n(x)\:,\en
  has not offered any solution to the eigenfunctions or eigenvalues. Very likely, the solutions are partially of the form $\psi_n(x)=(b^2-x^2)^{3/2}\,(remainder_n)$.
  Any further  information about the solutions of
  this differential equation would be welcome.
  
  \section{Extending The Particle in a Box\\ `Field Theory' that Leads to a \\Novel Version of Quantum Field Theory}
  Let us suppose that we accept AQ and its solution for a single item. Suppose we let $b$ become huge, and that we add terms to the potential so that the
  new quantum Hamiltonian (restoring the usual factor $1/2$) takes the form,
  \bn &&\H= \Sigma_{\bk}^{\bN} \half[\htp(\bk)^2 +\hbar^2(
  2\htv(\bk)^2+b^2)/(b^2-\htv(\bk)^2)^2
  + m^2\,\htv(\bk)^2 ] \no \\ &&\hskip19em
  +g\,\htv(\bk)^r \;\tau \:. \en
  Next, we add a new term, like gradients  $\Sigma_{\bk^*} (\tv(\bk^*) -\tv(\bk)^2/2\:a^2$, that represents a proper covariant term. Now, as $b\ra\infty$, it leads to a familiar form for the
  standard expression, which leads CQ to a grand summation formulation. 
  
  \subsubsection{ Some Monte Carlo results for the model $\vp^4_4$}
  We note that the field model, $\vp^4_4$, using  both CQ and  AQ, has been involved with several Monte Carlo 
  studies.\footnote{ Note that Monte Carlo, in
   affect, works a lot like a `grand summation'.} 
   
   The studies that used CQ 
   have led to  `unacceptable results', namely, as if the interaction term was absent when, in fact, it was active [4 - 7]. However, adopting AQ,  and using  the same Monte Carlo procedures, has  
   found  `acceptable results', by finding notable behavior from the interaction term [8, 9].
   
   Let us examine  that good behavior using an AQ formulation, and by introducing a novel procedure.
   
   \section{The Introduction of the `Anti-Box' }
   Our present `box' has been created by removing the coordinate points at $q=\pm b$, and then keeping the region $|q|<b$, while  discarding $|q|>b$. For our  `anti-box', the `only change made', is that now we {\it keep} $|q|>b$ and {\it discard} $|q|<b$. To deal with large $q$-values, we add potential terms, such as  $m^2 q^2$ and $g\,q^r$. 
   
   Jumping way ahead, and again restoring the 
   usual factor of $1/2$ to the Hamiltonian, we introduce the grand summation for the classical Hamiltonian, given
    (with AB $\Ra$ `Anti-Box') by
   \bn  H_{AB}=\Sigma_{\bk}^{\bN}\{\half \,[\tp(\bk)^2  +
  m^2\,\tv(\bk)^2 \,]  + g\,\tv(\bk)^r \,\}\:\tau\;,
  \en which becomes the CQ quantum Hamiltonian 
  \bn \H_{CQ-AB}= \Sigma_{\bk}^{\bN}\{\,\half \,[\,\htp(\bk)^2  + m^2\,\htv(\bk)^2\;] + g\,\htv(\bk)^r \,\}\:\tau\;.\en
  
  Now switching from CQ to  AQ, the AQ  quantum Hamiltonian is given by 
  \bn &&\H'_{AQ-AB}=\Sigma_{\bk}^{\bN}\{\,\half\,[\,\htp(\bk)^2 +\hbar^2\,(
  2\htv(\bk)^2+b^2)/(b^2-\htv(\bk)^2)^2 + \no \\
  &&\hskip10em 
  + m^2\,\htv(\bk)^2 \:] +g\,\htv(\bk)^r \,\}\:\tau\;.
  \en
  
  In this case, since we are in the `Anti-Box' realm, and where $0<b<\infty$, we do not let $b\ra\infty$ again, but  this time, we  let $b\ra0$, which leads to a nearly complete coordinate space for which  $-\infty<q\neq0<\infty$. This important fact leads to an AQ form of the quantization with  the new coordinate space, which is given (now using AB0 \,$\Ra $`Anti-Box-$b\ra0$') by
  \bn \H'_{AQ-AB0}=\Sigma_{\bk}^{\bN}\{\,\half\,[\,\htp(\bk)^2 +2\hbar^2/\htv(\bk)^2\,
  + m^2\,\htv(\bk)^2 \,] + g\,\htv(\bk)^r \,\}\:\tau\;.
  \label{77} \en
  
  \subsubsection{Rotational invariance}
  So far we have focused on  a single coordinate space rather than a multi-dimensional coordinate space. Our location of singular points can be realigned to independent linear paths that all cross over a fixed single point, which ultimately becomes the single point   
  that is removed from the entire space. As fully independent paths, the separate lines are made up from individual contributions.  In that case our AQ quantum Hamiltonian would become
\bn {\H''}_{AQ-AB0}=\Sigma_{\bk}^{\bN} \{ \,\half\,
[\, \ppp(\bk)^2+2\hbar^2/ \vvv (\bk)^2 + m^2\vvv(\bk)^2\,] + g \vvv(\bk)^r\,\}\;\tau\:.\en

The vector in this equation refers to the spatial coordinates, and it has multiple values in which we can let $\bk=(k_1, k_2, ..., k_s)$.
  
 \subsubsection{A note regarding two $\hbar$-factors in AQ}

  When the active coordinate space is only $ q>0$, it has been well determined \cite{lg} that the $\hbar$-term, i.e., the $(3/4)\hbar^2$ factor, is correct.
 However, we now require a larger linear coordinate space, where $|q|>0$, and there are both positive and negative contributions  present. Classical mechanics allows one to  just join the behavior of $q>0$ and $q<0$ to get the right physics for $|q|>.0$. However, that is definitely {\it not} good physics for quantum mechanics.
 
 All of the present AQ Monte Carlo studies have used the $(3/4)\hbar^2 $ factor while adopting both positive and negative field variables. Useful data was observed that proved that the AQ results were different from the CQ results. However, raising the factor term from $3/4$ to $2$, which now is proposed to correctly represent the  fact that positive and negative fields need a different $\hbar$-factor. Accepting the lager $\hbar$-factor can only enhance the AQ results!\footnote{The expression of eigenfunctions for `the particle in a box' are of the form  $\zeta_n(x)=(b^2-x^2)^{3/2}(remainder_n)$, which becomes, partially at least, an eigenfunction  for  (\ref{77}). Now, after $b\ra0$, this changes the previous expression to become  $\psi_n(x)=x^3(remainder'_n)$. This last expression enforces continuity for each eigenfunction and its first two derivatives in the region around $x=0$.}
 
 \section{Summary}
 We have adopted `The particle in  a box', a common toy model for beginners \cite{k1}, which is now  promoted to a `field theory' that is also focussed on a `Grand Summation' that contains a colossal, but finite, number of our `atom sized blocks' rather than pass to the continuum limit  of an integral. This formulation has been adopted after choosing a common size of atoms, which is truly tiny, but not zero. The procedure of skipping an integral to choose one of its colossal grand summations, has the virtue of none of the mathematical problems faced by conventual quantum field theory.  After all, our physical world is   composed of atoms and not by zero dimensional, infinitely many, mathematical points. As noted before, a grand summation has far  more physics pertaining to quantum theory than the aspects inferred by mathematical integration. 
 
 The question may be asked what about wave functions involving integrations and infinities. The coordinate $x$ is an item of the classical realm, and it has been welcomed to be used in quantum theory. The single operator pair $P$ and $Q$, where, $Q|x\>=x|x\>$, and for which $[Q,P]=i\hbar\,1\!\!1$,  offer wave functions, $\psi(x)=\<x|\psi\>$, provided that $\tint\,|\psi(x)|^2\;dx<\infty$. 
 
 Observe that we accept wave functions, such as $\psi(x) $, and do {\bf  NOT} treat them to our `multiple atom blocks' approach. This 
 is because wave function expressions, and their integrated results, are  {\it not} part of nature. Instead, they represent  {\it probabilistic  features}. This feature appears as  $\tint_A^B \,|\psi(x)|^2\;\ds$, which asserts that the location probability of their particular particle being found in the interval $B-A$, provided, of course, that $\tint_{-\infty}^\infty\;|\psi(x)|^2\;\ds=1$, which implies $100 \%$ certainty.

\end{document}